\documentstyle[11pt,aaspp4]{article}

\def\la{\mathrel{\hbox{\rlap{\hbox{\lower4pt\hbox{$\sim$}}}\hbox{$<$}}}}
\def\ga{\mathrel{\hbox{\rlap{\hbox{\lower4pt\hbox{$\sim$}}}\hbox{$>$}}}}

\slugcomment{Accepted to {\it The Astrophysical Journal}}

\begin{document}

\title{Gamma-Ray Burst Dust Echoes Revisited:  Expectations at Early Times}

\author{Jane A. Moran\altaffilmark{1} and Daniel E. Reichart\altaffilmark{1}}
\altaffiltext{1}{Department of Physics and Astronomy, University of North Carolina at Chapel Hill, Campus Box 3255, Chapel Hill, NC 27599; reichart@physics.unc.edu}

\begin{abstract}

Gamma-ray burst (GRB) dust echoes were first proposed as an alternative
explanation for the supernova-like (SN-like) components to the afterglows of
GRB 980326 and GRB 970228.  However, the spectroscopic identification of Type
Ic SN 2003dh associated with GRB 030329, as well as the identification of
SN-like components to the afterglows of other GRBs, appears to have confirmed
the GRB/SN paradigm.  However, the likely progenitors of Type Ic SNe are
Wolf-Rayet WC stars, and late-type WC stars have been observed to be
surrounded by dust, at a distance of $10^{14}$ -- $10^{15}$ cm from the star. 
Consequently, we revisit the possibility of GRB dust echoes, not on a
timescale of weeks after the burst but on a timescale of minutes to hours.  We
find that if the optical flash is sufficiently bright and the jet sufficiently
wide, GRB afterglows may be accompanied by chromatic variations on this
timescale.  From these signatures, such model parameters as the inner radius
of the dust distribution, the initial opening angle of the jet, etc., may be deduced.
 With rapid and regular localizations of GRBs by HETE-2, Integral, and now Swift, and
new and improved robotic telescope systems, these early-time GRB dust echoes
may soon be detected.  We describe one such robotic telescope system, called
PROMPT, that the University of North Carolina at Chapel Hill is building at
the Cerro Tololo Inter-American Observatory in greater detail.

\end{abstract}

\keywords{dust, extinction --- gamma rays: bursts --- stars: mass loss --- stars: Wolf-Rayet --- supernovae: general --- telescopes}

\section{Introduction}

GRB dust echoes were first proposed by Waxman \& Draine (2000) and Esin \&
Blandford (2000) as an alternative explanation for the SN-like components to
the afterglows of GRB 980326 (Bloom et al. 1999) and GRB 970228 (Reichart
1999; Galama et al 2000).  Here, ``dust echo'' refers to light that is either
(1) scattered by dust into the line of sight or (2) absorbed by dust and then
thermally reemitted into the line of sight.  In both cases, light is delayed
due to the greater path lengths.  Reichart (2001) modeled and computed dust
echo light curves and spectral flux distributions (SFDs) for both cases and
found that while dust echoes can mimic SN light curves they cannot mimic SN
SFDs, at least not near the spectral peak:  For example, the second component
to the afterglow of GRB 970228 cannot be explained by a dust echo because its
spectral peak is too narrow for the first case and at too high of a frequency
for the second case.  

The spectroscopic identification of SN 2003dh associated with GRB 030329
(Stanek et al. 2003), as well as the spectroscopic identification of SN 1998bw
associated with the unusual GRB 980425 (Galama et al. 1998) and the
identification of SN-like components to the afterglows of at least seven other
GRBs (e.g., Bloom et al. 1999; Reichart 1999; Galama et al. 2000; Bloom et al.
2002; Garnavich et al. 2003; Price et al. 2003; Della Valle et al. 2003; Cobb et al. 2004; Levan
et al. 2005), appears to have confirmed the GRB/SN paradigm,
at least for most long-duration/soft-spectrum GRBs. 

The leading models for making GRBs from collapsing massive stars are the
collapsar model (e.g., Woosley 1993) and the supranova model (e.g., Vietri \&
Stella 1999), both of which result in the formation of a black hole and an
accretion disk from which an ultra-relativistic jet is produced, but in the
case of the supranova model a neutron star that must first shed angular
momentum forms first.  In the case of the collapsar model, the progenitor must
first shed its outer layers in a Wolf-Rayet (WR) wind (or be stripped of them
by a companion) if the jet is to escape the progenitor.  Indeed, both SN
2003dh and SN 1998bw are Type Ic SNe, the progenitors of which have lost both
their hydrogen and helium layers.

Given that the SN start time matches the GRB time to within about two days for 
both of these events (e.g., Iwamoto et al. 1998; Hjorth et al. 2003),
the collapsar model is favored, at least
for these events.  However, given that there is no overlap between the sample
of GRBs for which SNe and SN-like components to afterglows have been detected
and the sample of GRBs for which blueshifted soft X-ray lines have been
detected (e.g., Piro et al. 2000; Reeves et al. 2002, 2003; Butler et al.
2003; Watson et al. 2003) -- which can be interpreted in terms of the supranova
model -- a dichotomous scenario cannot yet be ruled out.  However, in this
paper we will narrow our focus and consider only GRBs with WC (or WO)
progenitors, the most evolved of the WR stars.

Given that late-type WC stars are expected to be surrounded by dust at a
distance of $>$$10^{13}$ -- $10^{14}$ cm from the star and have been observed to be
surrounded by dust at a distance of $10^{14}$ -- $10^{15}$ cm -- not at a distance $\sim$$10^{18}$
cm as would be required to mimic a SN light curve -- we now revisit the
possibility of GRB dust echoes, not on a timescale of weeks after the burst
but on a timescale of minutes to hours.  In \S2, we summarize both the
expectation for and observations of dust at these distances from late-type WC
stars.  In \S3, we present model dust echo light curves and color histories for
a range of parameter values and discuss limitations of the model.  In \S4, we
draw conclusions and summarize a robotic telescope system that should be ideal
for identifying and studying dust echoes at early times after the burst.

\section{Dust around Late-Type WC Stars}

Many studies of infrared emission from late-type WC stars have shown them to
be dust-making machines (e.g., Allen, Harvey \& Swings 1972; Gehrz \& Hackwell
1974; Cohen, Barlow \& Kuhi 1975; Williams, van der Hucht \& Th\'e 1987). 
However, grains cannot condense out of the dense stellar wind within the
star's dust sublimation radius $r_s$.  For an optically thin medium in radiative
equilibrium:
\begin{equation}
4\pi a^2\overline{Q(a,T_s)}\sigma T_s^4 =
\pi a^2\overline{Q(a,T_\star)}\frac{4\pi R_\star^2\sigma T_\star^4}{4\pi r_s^2},
\end{equation}
where $\pi a^2 \overline{Q(a,T)}$ is the mean absorption cross-section for a grain of size $a$
averaged over a thermal spectrum of temperature $T$, $T_s$ is the sublimation
temperature, and $T_\star$ and $R_\star$ are the effective temperature and radius of the
star.  Taking $\overline{Q(a,T_s)} \approx \overline{Q(a,T_\star)}$ (e.g., Tuthill, Monnier \& Danchi
2001),
simplification yields: 
\begin{equation}
r_s \approx \frac{1}{2}\left(\frac{T_\star}{T_s}\right)^2R_\star.
\end{equation}
For an effective temperature $T_\star \sim 35$,000 K and radius $R_\star \sim 2 \times 10^{11}$ cm (e.g.,
Crowther 1997), and a sublimation temperature $T_s \sim 1$,500 K (e.g., Tuthill, Monnier \& Danchi
2001), the
sublimation radius is $r_s \sim 5 \times 10^{13}$ cm.  This is of course only a rough
estimate:  There is much ambiguity in how one measures the effective
temperature and radius of a WR star because of the opacity of its wind, and
the sublimation temperature varies with grain composition.  Furthermore, the
process by which dust forms beyond the sublimation radius is not fully
understood:  Dust is not expected to be present in large quantities at this
innermost possible radius.

Observations of dust around late-type WC stars yield distances that are
consistent with this expectation.  Williams, van der Hucht \& Th\'e (1987)
modeled infrared SFDs of 23 WC stars assuming an isotropic distribution of
dust and found inner radii that are typically in the range $10^{14}$ -- $10^{15}$ cm.  Others
have measured the dust distribution directly using one-dimensional speckle
interferometry (Allen, Barton \& Wallace 1981; Dyck, Simon \& Wolstencroft 1984
for WC9 star WR 104), aperture masking with Keck (Tuthill, Monnier \& Danchi
1999, 2002 for WR 104), lunar occultation (Ragland \& Richini 1999 for WC9 star
WR 112; Mondal \& Chandrasekhar 2002 for WR 104); and direct imaging with
HST-NICMOS2 (Marchenko, Moffat \& Grosdidier 1999 for WC7 star WR 137). 
However, all of these might be special cases:  WR 104 and WR 112 are the two
most extreme dust producers in the sample of Williams, van der Hucht \& Th\'e
(1987), presumably due to wind-wind collisions with short-period companions,
which additionally result in a highly anisotropic dust distribution; and WR
137, which is technically not a late-type WC star, is an episodic dust
producer due to wind-wind collisions with a long-period companion on an
elliptical orbit.  Despite these caveats, typical distances to their dust
distributions are $4 \times 10^{14}$ cm for WR 104, $6 \times 10^{14}$ cm for WR 112, and $7 \times 10^{15}$
cm for the less relevant WR 137.

\section{GRB Dust Echo Model and Limitations}

We now present dust echo light curves and color histories using the model of Reichart (2001), but for significantly smaller dust distributions than he explored.  We summarize the dust and light geometries and model parameters in the top panel of Figure 1:  Dust does not form within radius $R$ of the progenitor and is destroyed within half-angle $\theta_{jet}$ of the jet axis by bipolar jets (e.g., Waxman \& Draine 2000; Fruchter, Krolik \& Rhoads 2001).  Exterior to the solid curve, the dust density $n(r) \propto r^{-2}$.  The optical depth through this dust distribution is $\tau_{\nu} = n(R)\sigma_{\nu}R$, where $\sigma_{\nu}$ is the total (absorption plus scattering) cross section of the dust grains at frequency $\nu$.  A pulse of light is assumed to be emitted within half-angle $\theta_{col}$ of the jet axis (Reichart 2001).  

Light that is emitted at angle $\theta$ relative to the line of sight, where $\theta_{jet} < \theta < \theta_{col}$, and is scattered by dust at radius $r > R$ into the line of sight will arrive at the observer with an observer-frame time delay given by:
\begin{equation}
t = \frac{r(1+z)}{c}(1 - \cos\,\theta).
\end{equation}
In the bottom panel of Figure 1 we plot curves of constant arrival time for an on-axis jet.  The spectral flux of the dust echo at time $t$ is then given by integrating along the illuminated portions of the corresponding curve the product of (1) the spectral flux of the pulse of light, (2) the dust density $n(r,\theta)$, (3) the probability that the light will scatter an angle $\theta$ into the line of sight, and (4) $e^{-\tau_{\nu (1+z)}(r,\theta)}$, where $\tau_{\nu (1+z)}(r,\theta)$ is the optical depth integrated along the light path.  Expressions for the differential scattering cross section and $\tau_{\nu (1+z)}(r,\theta)$, as well as useful analytic approximations for the light curve and color history of the dust echo, can be found in \S2.1 of Reichart (2001).

In Figures 2 -- 5, we plot dust echo light curves and color histories in which
we vary each of the four model parameters -- $R$, $\theta_{jet}$, $\tau_{\nu (1+z)}$, and $\theta_{col}$ -- while holding the other three constant:

\noindent {\it Figure 2:}  Greater values of $R$ result in later turn-on times because of the greater path length that the light must travel.  Greater values of $z$ result in later turn-on times because of cosmological time dilation.  The dust echo starts out bluer than the optical flash because blue light scatters preferentially, but grows redder than the optical flash because of increasing path lengths through dust.  The re-brightening of the dust echo at later times corresponds to light from the jet that is pointed away from us backscattering off of the far side of the dust distribution.  This light is bluer because destruction of dust by the two jets leaves little dust in the path of this light to redden it.  

\noindent {\it Figure 3:}  Greater values of $\theta_{jet}$ result in later turn-on times because of the greater path length that the light must travel.  The re-brightening of the dust echo at later times occurs an equal amount of time earlier. 

\noindent {\it Figure 4:}  Greater values of $\tau_{\nu (1+z)}$ result in dust echoes that grow redder because of increasing path lengths through greater dust densities. 

\noindent {\it Figure 5:}  Greater values of $\theta_{col}$ result in dust echoes that are brighter at intermediate times because more light is available to be scattered by dust. 

The smaller values of $R$ that we consider in this paper permit two
simplifications:  (1) We have assumed that dust within $\theta_{jet}$ is destroyed to well beyond $R$ (which is equivalent to setting $f = 1$ in Reichart 2001); and (2) We can limit the discussion to light scattered from the optical flash, which we take to be light from either internal shocks (e.g., Vestrand et al. 2005) or the reverse shock of the afterglow:  Reichart (2001) showed that light scattered from the forward shock of the afterglow can only outshine the afterglow for large values of $R$.

Consider the case of an optical flash of the form:
\begin{equation}
F_\nu (t) = \cases{F_{\nu ,0} & ($t < t_0$)\cr F_{\nu ,0}(t/t_0)^{-2} & ($t > t_0$)},
\end{equation}
where $F_{\nu ,0}$ is the brightness of the optical flash and $t_0$ is the timescale
of
the optical flash, after which it fades as $t^{-2}$.  The spectral fluence of the
optical flash is then $2F_{\nu ,0}t_0$ and by Equation 24 of Reichart (2001) the
peak
brightness of the dust echo is given by:
\begin{equation}
\frac{F_{\nu ,DE}}{F_{\nu ,0}} \approx 0.0045\left(\frac{\tau_{\nu(1+z)}}{1+z}\right)\left(\frac{R}{3 \times 10^{14}\,\,\rm{cm}}\right)^{-1}\left(\frac{t_0}{30\,\,\rm{ sec}}\right),
\end{equation}
where $F_{\nu ,DE}$ is the peak brightness of the dust echo,
$\tau_{\nu(1+z)}$ is the optical
depth through the dust distribution at frequency $\nu(1+z)$, and $z$ is the
redshift.\footnote{Equation 24 of Reichart (2001) has a factor of $e^{-\tau_{\nu(1+z)}}$ that we
have dropped
here, because Equation 24 is an approximation that does not hold around $t
\approx
t_{DE}$ when $\tau_{\nu(1+z)} \ga 1$.  However, from Figure 7 of Reichart (2001)
it is clear
that dropping this factor approximates the correct effect.
}  By Equation 5 of Reichart (2001), the peak time of the dust echo
is given by:
\begin{equation}
\frac{t_{DE}}{t_0} \approx 17\left(\frac{R(1+z)}{10^{15}\,\,\rm{cm}}\right)\left(\frac{\theta_{jet}}{10^\circ}\right)^2\left(\frac{t_0}{30\,\,\rm{ sec}}\right)^{-1},
\end{equation}
where $t_{DE}$ is the peak time of the dust echo.  For these parameter values,
$\log(F_{\nu ,DE}/F_{\nu ,0})/\log(t_{DE}/t_0) \approx -1.9$ and consequently
the dust echo would
outshine the optical flash beginning at $t_{DE} \approx 8$ min, which would result
in a
$\approx$0.9 mag bump in the light curve at this time, if the forward shock is
not yet
a major contaminant.  Furthermore, the dust echo would be even more prominent
at bluer wavelengths, since $\tau_{\nu (1+z)}$ is an increasing function of $\nu$.  Greater values of $\tau_{\nu(1+z)}$, $R$, $z$, and 
especially $\theta_{jet}$,
and lesser values of $t_0$, might also make the dust echo more prominent, in part
by delaying it.  However, the forward shock is more of a contaminant at later
times as well.  Even so, for $\tau_{\nu(1+z)}$, $R$, and $\theta_{jet}$ three
times larger and $t_0$ 
three times smaller, $\log(F_{\nu ,DE}/F_{\nu ,0})/\log(t_{DE}/t_0) \approx
-0.9$, which means that the
dust echo would outshine the forward shock beginning at $t_{DE} \approx 11$ hr as
long as
the optical flash outshines the forward shock at $t_0 \approx 10$ sec (for a
forward
shock that fades faster than $t^{-0.9}$).

However, these estimates for the scattering of light by grains of dust must
also be taken with a few grains of salt.  The model of Reichart (2001) makes a
number of limiting assumptions:  (1) The viewing angle is assumed to be zero;
(2) Light that scatters is assumed to not scatter again into the line of sight; and (3) The radius at which the optical flash occurs is assumed to be much less than $R$.

The effect of a non-zero viewing angle will be to stretch out the turn-on time
of the dust echo, from $t \approx 0.1t_{DE}$ to a delayed peak time of $t \approx 
2.7t_{DE}$ for the
typical viewing angle (Reichart 2001).  The light curve and color history of
this transition is difficult to predict without performing numerical
simulations.  The scattering assumption should be fine for
$\tau_{\nu(1+z)} < 1$ but
will become increasingly less accurate at greater optical depths. 

Although internal-shock optical flashes should occur at smaller radii and not be a problem, reverse-shock optical flashes should begin at the deceleration radius, which
for a wind-swept environment is given by:
\begin{equation}
r_d = \frac{\eta {\cal E}v_w}{c^2\dot{M}\Gamma_0^2},
\end{equation}
where $\eta$ is a constant of proportionality that is equal to $9/2$ in the
case of
the Blandford-McKee solution, $\cal E$ is the isotropic-equivalent energy release of
the explosion, $v_w$ is the velocity of the progenitor's wind, $\dot{M}$ is the
mass-loss rate of the progenitor, and $\Gamma_0$ is the initial Lorentz factor
of the
ejecta (e.g., Kumar \& Panaitescu 2003).  For $\eta = 9/2$, ${\cal E} = 10^{52}$ erg, $v_w =
2$,000
km/s, $\dot{M} = 5 \times 10^{-5}$ M$_\odot$/yr, and $\Gamma_0 = 300$, $r_d = 3 \times 10^{13}$ cm.  Since
this is
considerably less than the values of $R$ that Reichart (2001) considered, he was
able to safely ignore this offset from the center of the dust distribution in
his model, which assumes spherical symmetry (except along the jet axis where
dust is destroyed).  However, now this case is marginal:  For larger values of
$\cal E$ in particular, this model will no longer apply and light curves and color
histories will be difficult to predict without performing numerical
simulations.  However, given the standard-energy result (e.g., Frail et al.
2001; Bloom et al. 2003) larger values of $\cal E$ imply smaller values of
$\theta_{jet}$, the
dust echoes for which would be difficult to detect anyway (Equation 6).

However, the model of Reichart (2001) might still apply to reverse-shock optical flashes, even for large values
of $\cal E$.  For example, if the central engine is long lived only a fraction of
this energy might be released initially, in which case the optical flash would
occur at a smaller distance.  Also, $R$ might be greater than what we expect
from late-type WC stars:  If the progenitor's final years are accompanied by
greater ultraviolet output, dust will be sublimated to greater distances. 
Furthermore, greater charging of carbon atoms might prevent grain formation
altogether (e.g., Williams, van der Hucht \& Th\'e 1987), in which case $R$ would
be set by the wind velocity and the time since last significant grain
formation, possibly increasing $R$ by a factor of ten or more.

\section{Conclusions and PROMPT}

We have revisited the possibility of GRB dust echoes in light of expectations
for a WC star progenitor.  We find that if the optical flash is sufficiently
bright and the jet sufficiently wide, GRB afterglows may be accompanied by
detectable chromatic variations, not on a timescale of weeks after the burst
but on a timescale of minutes to hours.  From these signatures, such model
parameters as the inner radius of the dust distribution, the initial opening angle of
the jet, etc., may be deduced.

However, in order for early-time dust echoes to be identified and studied,
three conditions must first be satisfied:  (1) GRBs must be localized rapidly,
preferably within tens of seconds; (2) Optical/near-infrared telescopes must
respond on a similar timescale; and (3) These telescopes must observe in many
wavelength bands and preferably simultaneously, else it may be impossible to
disentangle SFDs from temporal variability, particularly at early times. 
Condition (1) is being satisfied by HETE-2, Integral, and now Swift, about once every fourth
day.  Condition (2) is being satisfied by a wide variety of
robotic telescopes.

The Panchromatic Robotic Optical Monitoring and Polarimetry Telescopes
(PROMPT) that the University of North Carolina at Chapel Hill is building at
the Cerro Tololo Inter-American Observatory will additionally satisfy
condition (3).  When completed in late-2005, PROMPT will consist of six
0.41-meter Ritchey-Chrétien telescopes on rapidly slewing (9$^\circ$/sec)
Paramount
ME mounts and will image GRB afterglows in nine wavelength bands -- 
u'g'r'Ri'z'YJH -- six of them simultaneously. 
(The R telescope will
additionally measure the polarization of the afterglow.)  In addition to
measuring redshifts by dropout and extinction curves in great detail,
especially when combined with Swift UVOT and XRT measurements, PROMPT's rapid
slewing and simultaneous multi-wavelength imaging design will make it ideal
for identifying and studying dust echoes at early times.

\acknowledgements
DER very gratefully acknowledges support from NSF's MRI, CAREER, PREST, and REU programs,
NASA's APRA, Swift GI, and IDEAS programs, Dudley Observatory's Ernest F. Fullam
Award, and especially Leonard Goodman and Henry Cox.

\clearpage

\figcaption[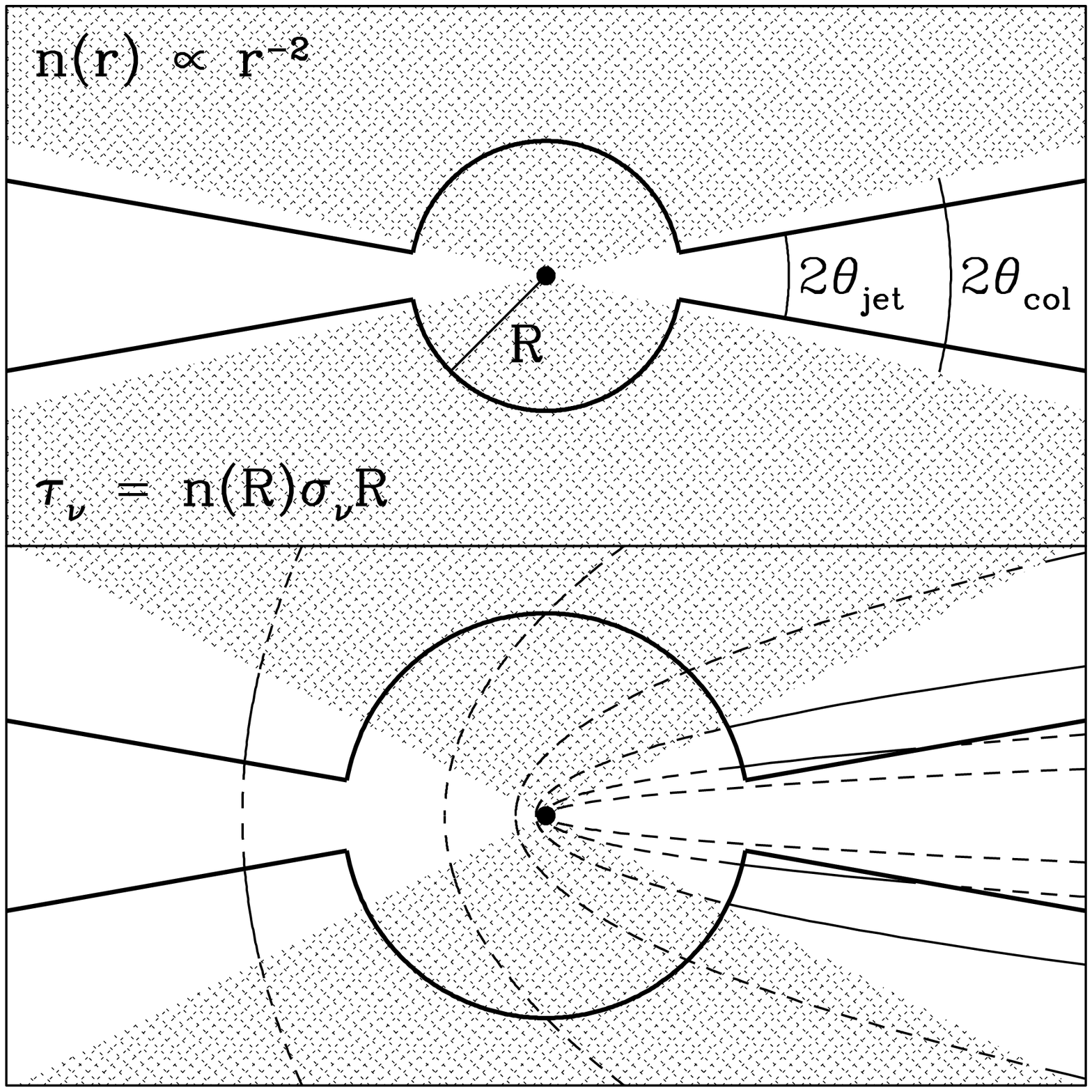]{{\it Top panel:}  Dust and light geometries and model parameters.  Dust does not form within radius $R$ of
the progenitor and is destroyed within half-angle $\theta_{jet}$ of the jet axis
by
bipolar jets (e.g., Waxman \& Draine 2000; Fruchter, Krolik \& Rhoads 2001). 
Exterior to the solid curve, the dust density $n(r) \propto
r^{-2}$.  The optical depth through this dust distribution is $\tau_{\nu} =
n(R)\sigma_{\nu}R$, where
$\sigma_{\nu}$ is the total (absorption plus scattering) cross section of the
dust grains
at frequency $\nu$.  A pulse of light is assumed to be emitted within
half-angle
$\theta_{col}$ of the jet axis (Reichart 2001).  {\it Bottom panel:}  Curves of constant arrival time $t = 0.01$, 0.03, 0.1, 0.3, 1 and 3 in units of $R(1+z)/c$.  Solid portions mark illuminated dust.}

\figcaption[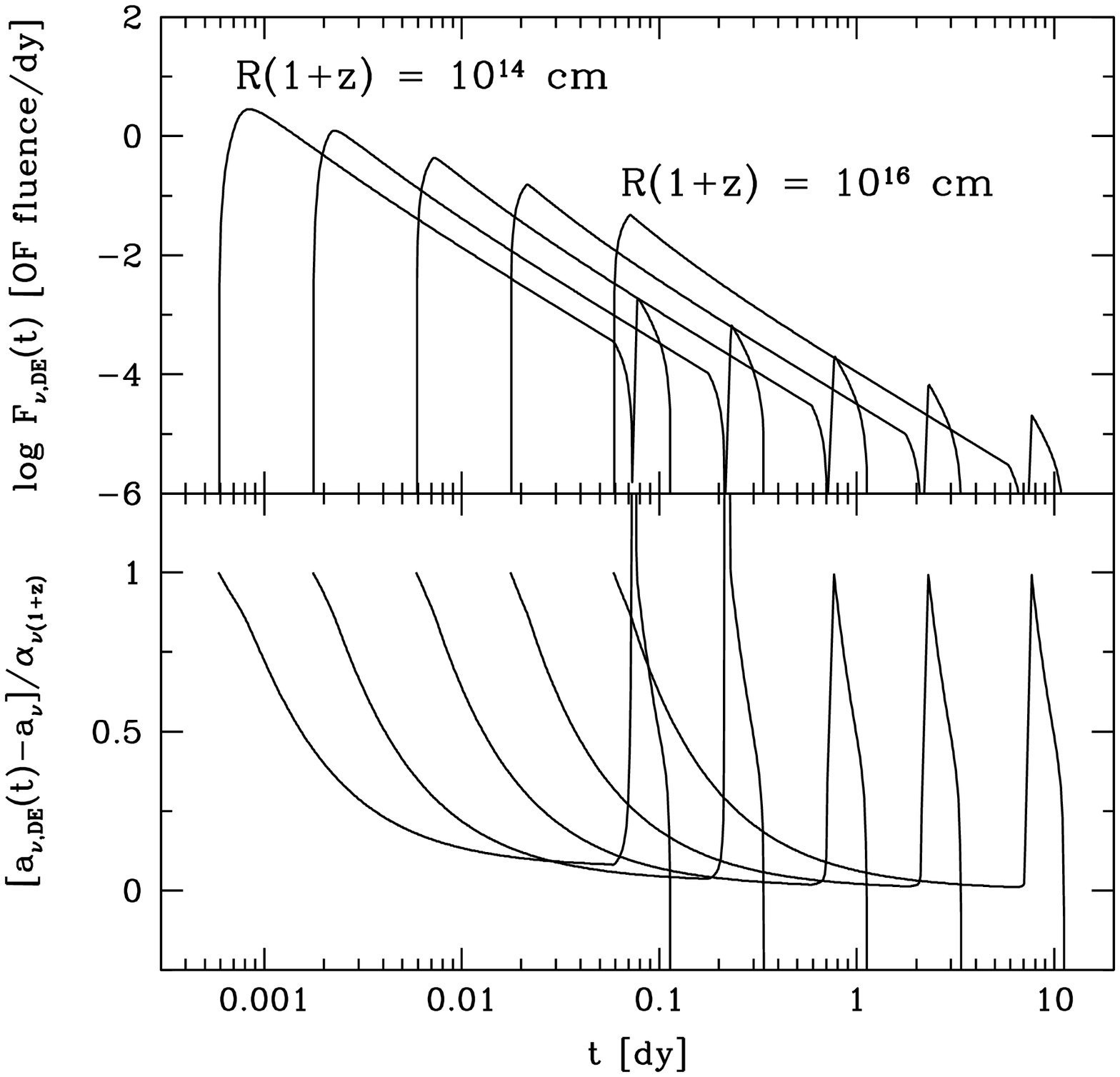]{Light curves (top panel) and color histories (bottom panel) of dust
echoes for $R(1+z) = 10^{14}$, $3 \times 10^{14,}$ $10^{15}$, $3 \times 10^{15}$, and $10^{16}$ cm, $\theta_{jet} =
10^\circ$,
$\tau_{\nu (1+z)} = 1$, and $\theta_{col} = 1.1\theta_{jet}$.  The light curves are normalized to the fluence of the optical flash divided by one day.  The parameter $a_{\nu,DE}(t)$ is the spectral
index
of the dust echo, $a_{\nu}$ is the spectral index of the optical flash, and $\alpha_{\nu (1+z)}$ is given by $\tau_{\nu (1+z)} \propto [{\nu (1+z)}]^{\alpha_{\nu (1+z)}}$.  The parameter $a_{\nu}$ is likely $1/3$, $-1/2$, or $-(p-1)/2$, where $p$ is the power-law index of the electron-energy distribution (e.g., Sari, Piran \& Narayan 1998), and $1 \la \alpha_{\nu (1+z)} \la 1.6$
(Reichart 2001).}

\figcaption[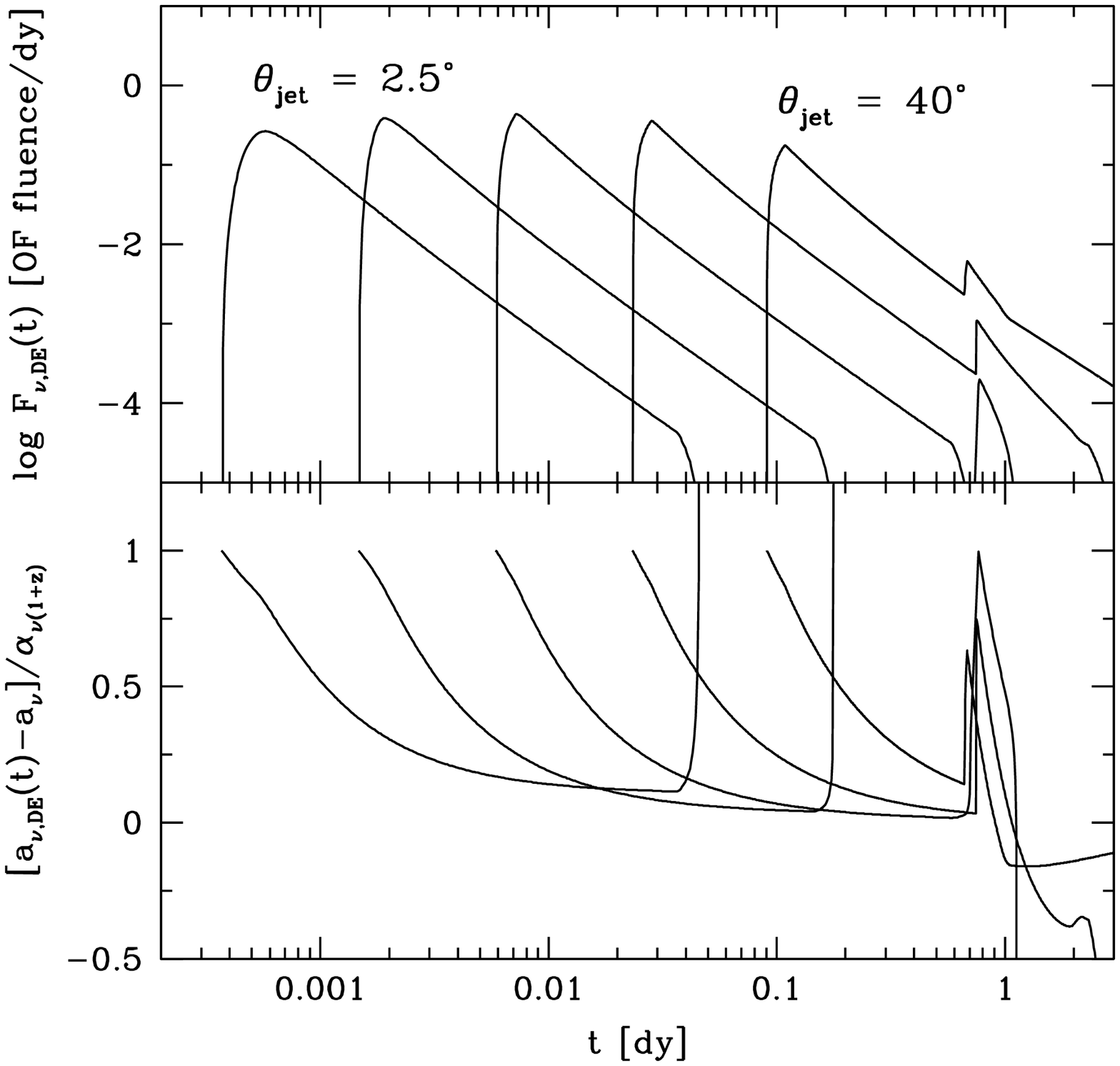]{Light curves (top panel) and color histories (bottom panel) of dust
echoes for $\theta_{jet} = 2.5^\circ$, $5^\circ$, $10^\circ$, $20^\circ$, and $40^\circ$,
$R(1+z) = 10^{15}$ cm, $\tau_{\nu (1+z)} = 1$,
and $\theta_{col} = 1.1\theta_{jet}$.}  

\figcaption[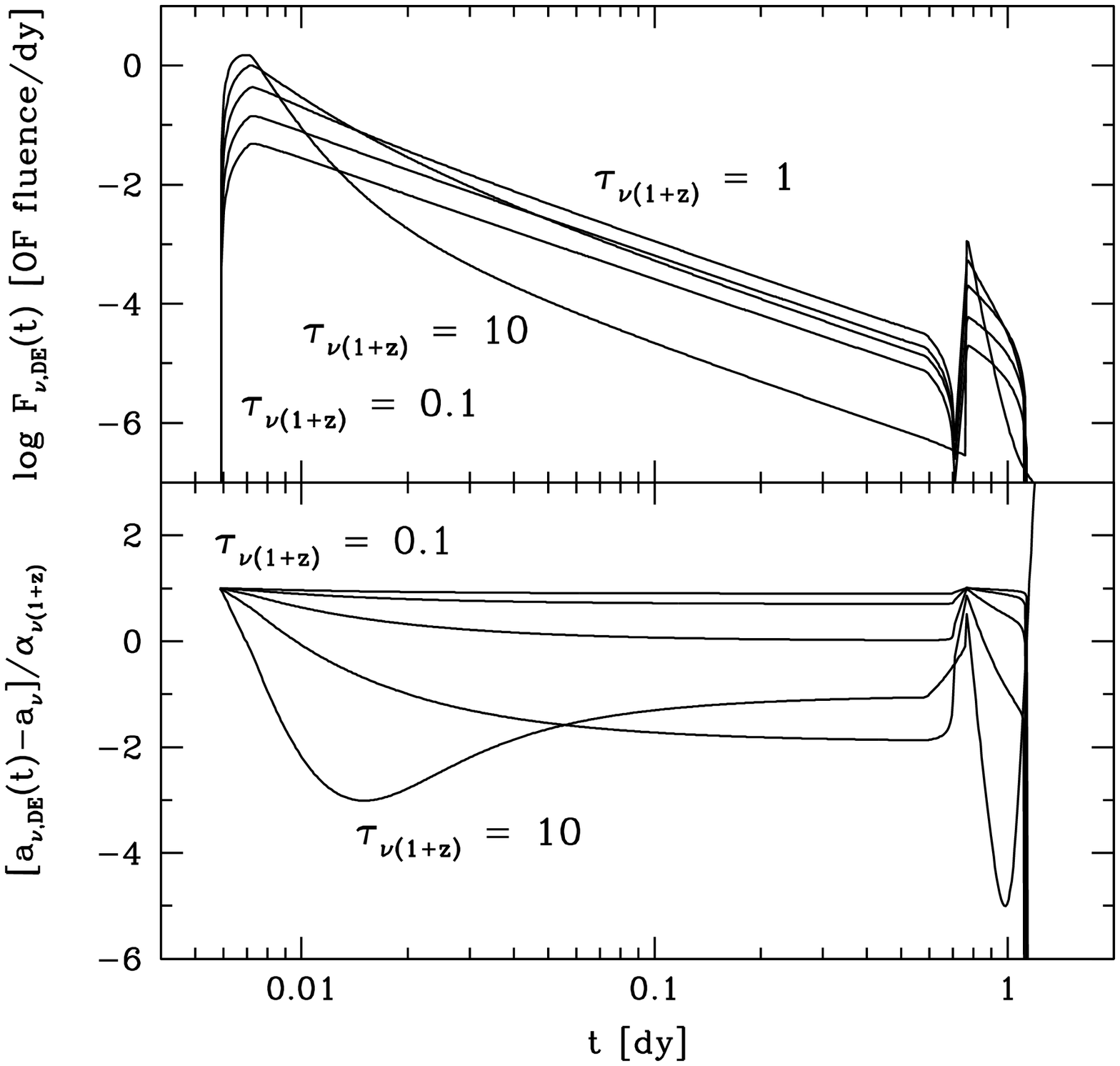]{Light curves (top panel) and color histories (bottom panel) of dust
echoes for $\tau_{\nu (1+z)} = 0.1$, 0.3, 1, 3, and 10, and $R(1+z) = 10^{15}$
cm, $\theta_{jet} = 10^\circ$,
and $\theta_{col} = 1.1\theta_{jet}$.}

\figcaption[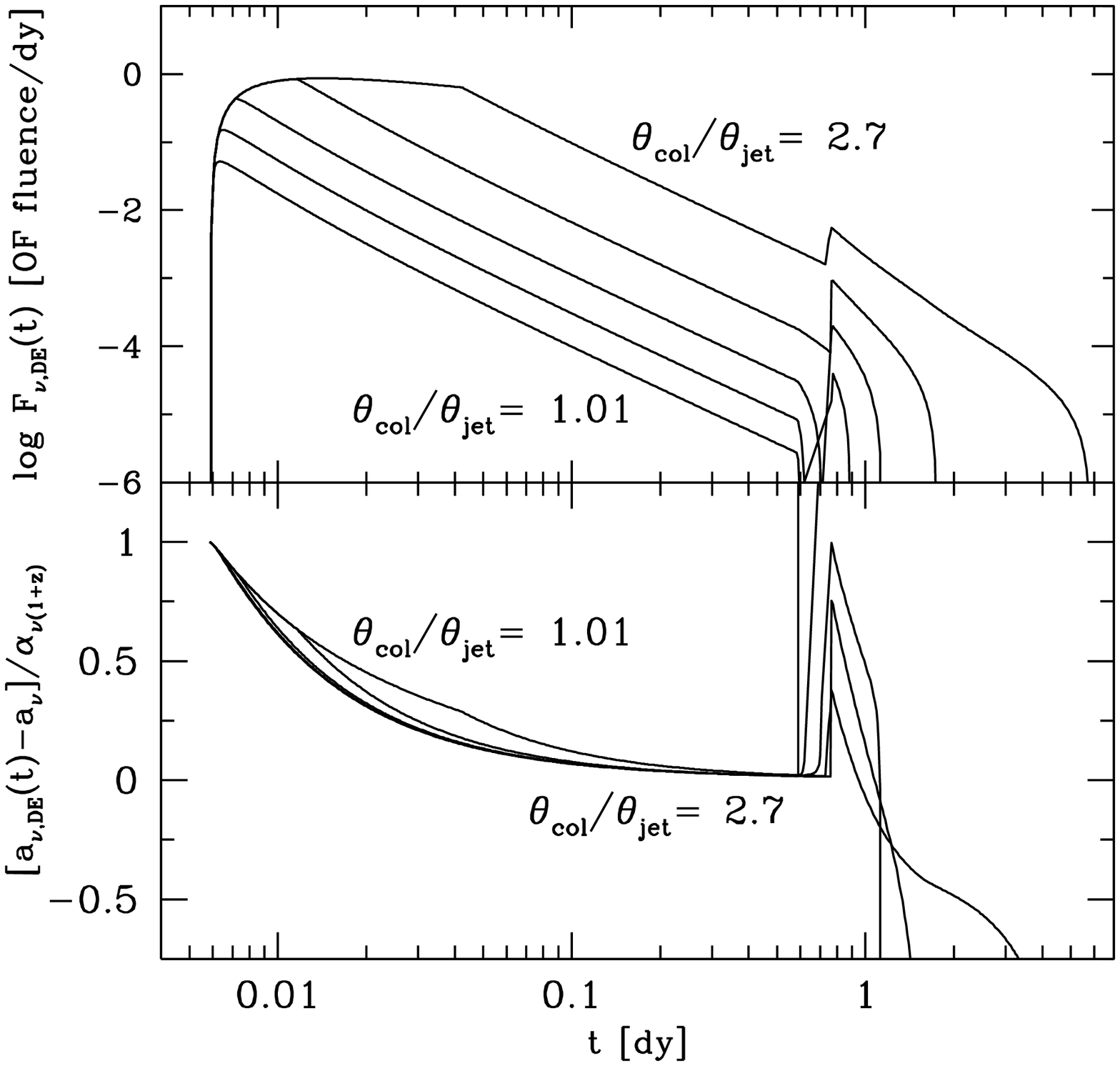]{Light curves (top panel) and color histories (bottom panel) of dust
echoes for $\theta_{col}/\theta_{jet} = 1.01$, 1.03, 1.1, 1.4, and 2.7, $R(1+z) =
10^{15}$ cm, $\theta_{jet} =
10^\circ$, and $\tau_{\nu (1+z)} = 1$.}

\clearpage

\setcounter{figure}{0}

\begin{figure}[tb]
\plotone{f1.eps}
\end{figure}

\begin{figure}[tb]
\plotone{f2.eps}
\end{figure}

\begin{figure}[tb]
\plotone{f3.eps}
\end{figure}

\begin{figure}[tb]
\plotone{f4.eps}
\end{figure}

\begin{figure}[tb]
\plotone{f5.eps}
\end{figure}

\end{document}